\documentclass[12pt]{article}
\newcommand{\beq}{\begin{equation}}
\newcommand{\eeq}{\end{equation}}
\newcommand{\bra}{\begin{array}}
\newcommand{\era}{\end{array}}
\newcommand{\be}{\beta}
\newcommand{\te}{\theta}
\newcommand{\al}{\alpha}
\newcommand{\ga}{\gamma}
\newcommand{\de}{\delta}
\newcommand{\da}{\dagger}

\newcommand{\om}{\omega}

\newcommand{\si}{\sigma}

\usepackage{latexsym}
\author{Jamila Douari \footnote{douari@sun.ac.za} \\
\small\it Stellenbosch Institute for Advanced Study, Private Bag X1,\rm\\
\small\it Matieland, Stellenbosch, 7601, South Africa\rm }
\title{Quasi-Particles Hamiltonian}
\begin{document}
\maketitle
\vspace*{0.5cm}

\section*{Abstract}
\hspace{.3in}The anyonic Hamiltonian is quantum mechanically given and the bosonic and the fermionic Hamiltonians are found as extremes by discussing the cases of the statistical parameter $\nu$ and the dimension of space. The anyonic algebra \cite{upa} is recalled as a deformed Heisenberg algebra and a deformed $C_\lambda$-extended Heisenberg algebra.

\section{Introduction}
\hspace{.3in}The important result we got in the reference \cite{upa} is a symmetry describing anyons called quasi-particles \cite{lm} which could be seen as a unified symmetry based on  the redefinition of the fundamental algebra underlying the non-commutative geometry \cite{bfs,sc,c}, and the redefinition of annihilation and creation operators by imposing the existence of an excitation operator. In the particular case of statistical parameter $\nu\in[0,1]$, we gave the anyonic algebra as a deformed $C_\lambda$-extended Heisenberg algebra, in which the Heisenberg algebra is extended by a polynomial in a hermitian operator denoted $K_i$ and deformed in terms of statistical parameter $\nu$. The obtained algebra is a deformed version of the extended Heisenberg algebra, discussed in the litterature \cite{ok,bi,pl,ply,bem}.\\

The considerable interest to the quasi-particles living in 2-dimensional (2d) space with fractional spin, charge and statistics \cite{lm} is conditioned nowdays by their applications to the theory of planar physical phenomena such as fractional quntum Hall effect, high-$T_{c}$ superconductivity \cite{mm} and quantum computers \cite{moc}. One of the approaches describing anyons and realizing a quantization of its theory to reveal fractional statistics is the group-theoretical approach analogously to the case of integer and half-integer spin fields.\\

In this letter, we recall the anyonic algebra and we give its representation. Then we extend the commutation relation between the annihilation and creation operators for higher orders which are useful for other results. Consequently we determine the corresponding Hamiltonian. This latter could be seen as a unified form for $d$-dimensional particles system in quantum mechanics. It is still under discussion! Thus, for $d\ge3$, the Hamiltonian is the elementary particles (bosons or fermions) Hamiltonian and if $d=2$ the Hamiltonian is the anyonic one.\\

The letter is organized as follows; In section 2, we review the anyonic algebra as a deformed Heisenberg algebra. In section 3, we recall the deformed $C_\lambda$-extended Heisenberg algebra describing anyons. In section 4, we extend the commutation relation between the annihilation and creation operators for higher orders. In section 5, we give the representation of anyonic algebra and the anyonic Hamiltonian. The section 6 is devoted for conclusion and remarks. 
\section{Anyonic Algebra}
\hspace{.3in}In the reference \cite{upa}, we suggested a new re-defnition of non-commutative geometry based on the statistical properties of space. It is defined by the coordinates $x_i$ and the momentum $p_i$ satisfying
\beq
\bra{rl}
\lbrack x_i ,x_j \rbrack_\chi = i\theta \epsilon_{ij}& ,\lbrack p_i ,p_j \rbrack_\chi = i\theta (\mu\om)^{2}\epsilon_{ij}
\era
\eeq
and for $i\le j$ we have
\beq
\bra{rl}
\lbrack x_i ,p_j \rbrack_\chi =i\delta_{ij}& ,\lbrack p_i ,x_j \rbrack_\chi = -i\delta_{ij}
\era
\eeq
with
\beq
\bra{rcl}
\chi=e^{\pm i\nu\pi},& \epsilon_{12}=-\epsilon_{21}=1,& \epsilon_{ii}=0,
\era
\eeq
$\nu\in\Re$ is a statistical parameter and $\pm$ sign indicates the two rotation directions on 2d space. $\te$ is a non-commutative real parameter satisfying the condition $$\te=(\frac{1}{\mu\om})^2\te^{-1}$$ which is necessary to maintain the bosonic statistics when $d\ge3$ and $\nu=0$.\\

Now, we suggest that, in 2d space, the particles are excited and they satisfy exotic statistics. So, we consider a unitary operator $\xi_i$ as excitation operator commuting with the spatial coordinates $x_{j}$, $\forall j$. In this case, by considering 2d harmonic oscillator which can be decomposed into 1d oscillators, so for each dimension we define the representation of the annihilation and the creation operators on 2d space as excited operators
\beq
\bra{ll}
b_{i}&= \sqrt{\frac{\mu\om}{2}}(x_i +\frac{i}{\mu\om} \xi_i p_i ) \\ \\
b^{\da}_{i}&= \sqrt{\frac{\mu\om}{2}}(x_i -\frac{i}{\mu\om} \xi^{-1}_i p_i ) .
\era
\eeq
with
\beq
\xi_i =e^{i\nu\pi K_i },
\eeq
and $K_i $ a hermitian operaor to be specified later.\\

We think of the suggested excitation as a phenomenon caused by the statistical properties of the non-commutative space, in which the particles are described by exotic statistics or called intermediate statistics \cite{lm}. This excitation will lead to a non-ordinary interaction. We call it a comparison interaction, statistical interaction or topological interaction since it is due to the dimension of space as we will see in the section 4.\\

Thus the algebra (1-2) of our non-commutative geometry leads to the following deformed Heisenberg algebra
\beq
\bra{ccc}
\lbrack b_{i},b^{\da}_{j} \rbrack_\chi &=&(\xi_i +\xi^{-1}_j )\de_{ij},\\ \\
\lbrack b^{\da}_{i},b^{\da}_{j} \rbrack_\chi &=& i\frac{\mu\om}{2}\te(I-\xi^{-1}_i \xi^{-1}_j ) \epsilon_{ij},\\ \\
\lbrack b_{i},b_{j} \rbrack_\chi &=& i\frac{\mu\om}{2}\te(I-\xi_i \xi_j )\epsilon_{ij},
\era
\eeq
with $I$ is the identity.\\

The algebra (6) is the Anyonic Algebra given in \cite{upa}. If we consider the space of dimension $d$ and if its peculiar cases deponding on its statistical properties we discussed in \cite{upa} are enough then the obtained algebra could be seen as a unified algebra. For $d=2$, the statistical parameter $\nu$ is arbitrary real parameter and the algebra describes anyonic system. If $d\ge 3$, $\nu=0,1$, the algebra describes bosons and fermions, respectively.\\
\section{A Deformed $C_{\lambda}$-Extended Heisenberg Algebra}
\hspace{.3in}In this section we recall the construction of anyonic algebra as a deformed $C_{\lambda}$-Extended Heisenberg Algebra.\\

The first commutation relation of the algebra (6) becomes
\beq
\lbrack b_{i},b^{\da}_{j} \rbrack_\chi = (I+\Re_{ij}^{[\nu]})\de_{ij},
\eeq
with $\nu\in[0,1]$ and by applying the Taylor expansion to $\xi_i$. The operator $\Re_{ij}^{[\nu]}$  is given by
\beq
\Re_{ij}^{[\nu]}=\sum\limits_{\ell=1}^{\frac{n+1}{2}}\kappa_{\nu,\ell}\frac{K_i ^{2\ell -1}-K_j ^{2\ell -1}}{2}+\sum\limits_{k=1}^{\frac{m}{2}}\kappa_{\nu,k}\frac{K_i ^{2k }+K_j ^{2k}}{2}
\eeq
with $m$ is even and $n$ odd, $n,m\in\bf N\rm$,  the powers of $K_i $ are odd in the first sum and even in the second and the $\kappa_{\nu,\ell}$ and $\kappa_{\nu,k}$ are given by the following expressions
$$\kappa_{\nu,\ell}=\frac{(i\nu\pi)^{2\ell -1}}{(2\ell -1)!},\phantom{~~~~}\kappa_{\nu,k}=\frac{(i\nu\pi)^{2k}}{(2k)!}.$$
By imposing $$K_i ^{\lambda} = I$$ for $\lambda\in\bf N\rm$ we have $n,m\le\lambda -1$.\\

Thus, we get the following commutation relations by the same method
\beq
\bra{ll}
\lbrack b^{\da}_{i},b^{\da}_{j} \rbrack_\chi = -i\frac{\mu\om}{2}\te\sum\limits_{\al=1}^{\lambda -1}\frac{(-i\nu\pi)^\al }{\al!}(K_i +K_j )^\al \epsilon_{ij},\\ \\
\lbrack b_{i},b_{j} \rbrack_\chi = -i\frac{\mu\om}{2}\te\sum\limits_{\al=1}^{\lambda -1}\frac{(i\nu\pi)^\al }{\al!}(K_i +K_j )^\al \epsilon_{ij}.
\era
\eeq
By introducing the defintion  $$K_i =e^{\frac{2\pi}{\lambda}N_i}$$ with $N_i$ is the number operator satisfying the following commutation relations
$$
\lbrack N_i, b_{j}\rbrack = -\de_{ij}b_{i}\phantom{~~~~~~~~~}\lbrack N_i, b^{\da}_{j}\rbrack = \de_{ij}b^{\da}_{i}$$
we get
\beq
K_i b_{j} = \de_{ij}e^{\frac{-2\pi}{\lambda}}b_{j}K_i \phantom{~~~~~~}K_i b^{\da}_{j} = \de_{ij}e^{\frac{2\pi}{\lambda}}b^{\da}_{j}K_i.
\eeq

We call the algebra (7-9-10) the quasi-particles algebra or anyonic one which is defined as a deformed $C_{\lambda}$-extended Heisenberg algebra, where $C_{\lambda}$ is a cyclic group $$C_{\lambda}=\{I, K_i , K^2_i ,K^3 _i ,...,K^{\lambda-1}_i \}.$$
\section{Extended Commutation Relations}
\hspace{.3in}Now let us extend the calculations for higher orders of annihilation and creation operators which will be useful for the representation of the anyonic algebra treated in the next section and also for other results; e.g. the construction of quasi-particles $w_\infty$ algebra \cite{wia}. For any order $\be$ of $b^{\da}_j$  the relation (7) is extended to the following commutation relation on non-commutative space
\beq
\lbrack b_{i},(b^{\da}_{j})^{\be} \rbrack_{\chi} = \Big ( \be +F_{ij}\Big ) \de_{ij} (b^{\da}_{i})^{\be -1},
\eeq
where
$$
F_{ij}=\sum\limits_{\ell=1}^{\frac{n+1}{2}}p_{\ell}\kappa_{\nu,\ell}\frac{K_i ^{2\ell -1}-K_j ^{2\ell -1}}{2}+\sum\limits_{k=1}^{\frac{m}{2}}p_{k}\kappa_{\nu,k}\frac{K_i ^{2k }+K_j ^{2k}}{2}
$$
with $p_{\ell}=\sum\limits_{\ga=0}^{\be-1}e^{\frac{-2(2\ell-1)\pi\ga}{\lambda}}$, $p_{k}=\sum\limits_{\ga=0}^{\be-1}e^{\frac{-2(2k)\pi\ga}{\lambda}}$,
and the powers of $K_i $ are odd in the first sum and even in the second.\\

Now, as a generalization of (11), for any order $\al$ and $\be$ of $b_i$ and $b^{\da}_j$ respectively we get
\beq
\lbrack b_i ^{\al},(b^{\da}_{j})^{\be}\rbrack_{\chi} = \sum\limits^{\al-1}_{k=0}\sum\limits^{k}_{\eta=0}\Big ( \be +F_{ij} q^{k}\Big ) G_{ij}^{[\eta]}(b^{\da}_{j})^{\be-\eta -1}(b_{i})^{\al-\eta -1},
\eeq
with $q=e^{\frac{-2\pi}{\lambda}}$ and according to (11) the operators $G_{ij}^{[\eta]}$ are defined by the following relation
\beq\bra{ll}
(b_{i})^{\al}(b^{\da}_{j})^{\be-1}&=G_{ij}^{[\al]}(b^{\da}_{j})^{\be-\al-1}+G_{ij}^{[\al-1]}(b^{\da}_{j})^{\be-\al}b_{i}+...\\ \\
&+G_{ij}^{[0]}(b^{\da}_{j})^{\be-1}(b_{i})^{\al},
\era\eeq
here we recite some of them, since the calculations are long and complicated for the orders $9<\al<\al-5$, so
\begin{displaymath}
\bra{llllll}
G_{ij}^{[0]}&=1\\
G_{ij}^{[1]}&=\al(\be-1)+\sum\limits_{\epsilon=0}^{\al-1}F_{ij}q^{\epsilon}\\
&.\\
&.\\
&.\\
G_{ij}^{[\al-1]}&=\prod\limits_{\sigma=1}^{\al-1}\Big ( (\be-\sigma)+F_{ij}q^{\al-\si}\Big ) +\\ \\
&+\sum\limits^{\al-2}_{\mu=2}\prod\limits_{\sigma=1}^{\al-\mu}\Big ( (\be-\sigma)+F_{ij}q^{\al-\si}\Big ) \prod\limits_{\sigma=\al-(\mu-1)}^{\al-1}\Big ( (\be-\sigma)+F_{ij}q^{\mu-\al-1}\Big ) \\ \\
&+\Big ( 2(\be-1)+\sum\limits^{\al-1}_{\eta=\al-2}F_{ij}q^{\eta}\Big ) \prod\limits_{\sigma=2}^{\al-1}\Big ( (\be-\sigma)+F_{ij}q^{\al-\si-1}\Big ) \\ \\
G_{ij}^{[\al]}&=\prod\limits_{\sigma=1}^{\al}\Big ( (\be-\sigma)+F_{ij}q^{\al-\si}\Big ) .
\era
\end{displaymath}

The commutation relations are used in \cite{wia} to construct the $w_\infty$ algebra characterizing the quasi-particles system living in non-commutative 2d space and to determine the states of Fock space in the following section. 

\section{Quasi-Particles Hamiltonian}
\hspace{.3in}Now, to find the Hamiltonian of the quasi-particles system we start by determining the number operator $N_i$ by using the above tools. According to (7), $N_i$ is given by
\beq
b^{\da}_{i}b_{i}=F(N_i )=\frac{1}{2-\chi}\Big (N_i+\sum\limits_{\mu=0}^{\frac{m}{2}}\sum\limits_{\ga=0}^{\mu-1}\kappa_{\nu,\ga}e^{\frac{4\pi\mu}{\lambda}(N_i)} \Big )
\eeq
and we can see that $$ b_{i}b^{\da}_{i}=F(N_i +1)=\frac{1}{2-\chi}\Big (N_i+1+\sum\limits_{\mu=0}^{\frac{m}{2}}\sum\limits_{\ga=0}^{\mu-1}\kappa_{\nu,\ga}e^{\frac{4\pi\mu}{\lambda}(N_i +1)} \Big ) $$
such that
$$\sum\limits_{\mu=0}^{\frac{m}{2}}\sum\limits_{\ga=0}^{\mu-1} e^{\frac{4\pi\mu}{\lambda}}\kappa_{\nu,\ga}=1$$
as a necessary condition to refind a fermionic number operator when $\nu=1$. The action of the number operator on the state $|n\rangle$ is given by
\beq
\bra{ll}
N_i |n\rangle =n|n\rangle ,& n\in\bf N\rm.
\era
\eeq
where the states are defined as follows
$$|n\rangle =\frac{1}{\sqrt{G_{ii}^{[n]}}}(b^{\da}_{i})^{n}|0\rangle $$
with $|0\rangle $ is the vacuum state and
$$G_{ii}^{[n]}=\prod\limits_{\sigma=1}^{n}\Big ( (n-\sigma+1)+F_{ii}q^{n-\si}\Big  ) $$
defined by (14). Thus, the following equalities
\beq
\bra{ll}
b^{\da}_{i}|n\rangle &=\frac{1}{2-\chi}\Big ( n+1+ \sum\limits_{\mu=0}^{\frac{m}{2}}\sum\limits_{\ga=0}^{\mu-1}\kappa_{\nu,\ga}e^{\frac{2\pi}{\lambda}(n+1)}\Big ) ^{\frac{1}{2}}|n+1\rangle \\ \\
b_i |n\rangle &=\frac{1}{2-\chi}\Big ( n +\sum\limits_{\mu=0}^{\frac{m}{2}}\sum\limits_{\ga=0}^{\mu-1}\kappa_{\nu,\ga}e^{\frac{2\pi}{\lambda}(n )}\Big ) ^{\frac{1}{2}}|n-1\rangle
\era
\eeq
leads to call $b_i$ the quasi-particles annihilation operator and $b^{\da}_{i}$ the quasi-particles creation operator.\\

By using the definition given in the bosonic Fock space representation as usual by
$$H_i=\frac{1}{2}\lbrace b_{i}, b^{\da}_{i}\rbrace,$$ and the above properties of the quasi-particles system, the associated Hamiltonian, is as follows
\beq
H_i =\frac{1}{2-\chi}\Big ( N_i +\frac{I}{2}+ \sum\limits_{\mu=0}^{\frac{m}{2}}\sum\limits_{\ga=0}^{\mu-1}\kappa_{\nu,\ga}K_i ^{2\mu} \frac{1+e^{\frac{4\pi\mu}{\lambda}}}{2}  \Big ).
\eeq

We remark that the obtained Hamiltonian is a deformed version of the ordinary one known for the elementary particles (bosons and fermions). The last term of the expression (17) could be treated as a statistical term which indicates the statistical interaction between the quasi-particles in the non-commutative 2d space. We also note that as extremes, in $d\ge 3$ case, we obtain bosonic Hamiltonian if $\nu=0$ and fermionic Hamiltonian if  $\nu=1$ such that the condition $\sum\limits_{\mu=0}^{\frac{m}{2}}\sum\limits_{\ga=0}^{\mu-1} e^{\frac{4\pi\mu}{\lambda}}\kappa_{\nu,\ga}=1$ is satisfied.\\
\section{Conclusion}
\hspace{.3in}To conclude, let us give an overview; First we introduced the anyonic algebra as interpolating symmetry beween bosonic and fermionic ones which are extremes of the obtained algbra. Then we got an open question about the possibility to consider the result as unified symmetry! In one of its peculiar cases, $d=2$ and $\nu\in[0,1]$, the algebra becomes a deformed $C_{\lambda}$-extended Heisenberg algebra describing anyons. By extending the calculations to higher orders, we determined the representation of anyonic algebra and then we gave the Hamiltonian as a deformed version of ordinary particles Hamiltonian with an extra term which could be seen as a statistical term. As remarks, if we look to the extremes of $\nu$ one notes that the Hamiltonian (17) becomes bosonic for $\nu=0$ and fermionic for $\nu=1$ in three or more dimensional space. Thus, let us ask the following questions. Could we imagine that the (17) has something to do with unified Hamiltonian describing $d$-dimensional Particles system for $d\ge2$? Is the last term of (17) identified with the statistical interaction indicated by the Chern-Simons term in the anyonic field theory?\\

Acknowledgements: The author would like to thank the Abdus Salam ICTP for the hospitality.

\end{document}